\documentclass[prb,showpacs,twocolumn,floatfix,superscriptaddress]{revtex4}

\usepackage{hyperref,fontenc,times,amsmath,amsfonts,amssymb,graphics,graphicx,epsfig,color,bbm,bm,units,mathtools,hhline}

\newcommand{\dd}{\mathrm{d}}
\newcommand{\ee}{\mathrm{e}}
\newcommand{\ii}{\mathrm{i}}
\DeclareMathOperator{\Tr}{Tr}
\DeclareMathOperator{\ad}{ad}

\DeclareMathOperator{\arcosh}{arcosh}
\DeclareMathOperator{\arsinh}{arsinh}
\DeclareMathOperator{\artanh}{artanh}
\DeclareMathOperator{\const}{const}
\DeclareMathOperator{\sinch}{sinch}
\DeclareMathOperator{\var}{var}
\newcommand{\super}[1]{^{(\!#1\!)}}

\DeclarePairedDelimiter\abs{\lvert}{\rvert}
\DeclarePairedDelimiter\bra{\langle}{\rvert}
\DeclarePairedDelimiter\ket{\lvert}{\rangle}
\DeclarePairedDelimiterX\braket[2]{\langle}{\rangle}{#1\delimsize\vert}
\DeclarePairedDelimiterX\braxket[3]{\langle}{\rangle}{%
  #1\,\delimsize\vert\,#2\,\delimsize\vert\,#3}

\newcommand{\brasub}[2]{\prescript{}{#2\!}{\bra{#1}}}
\newcommand{\ketsub}[2]{\ket{#1}_{\!#2}}

\makeatletter
\renewcommand*\env@cases[1][1.2]{%
  \let\@ifnextchar\new@ifnextchar
  \left\lbrace
  \def\arraystretch{#1}%
  \array{@{}l@{\quad}l@{}}%
}
\makeatother

\begin{document}
\title{True Limits to Precision via Unique Quantum Probe}
\author{Sergey Knysh} \email{Sergey.I.Knysh@nasa.gov}
\affiliation{Quantum Artificial Intelligence Laboratory (QuAIL), 
NASA Ames Research Center,  Moffett Field, California 94035, USA}
\affiliation{SGT Inc., 7701 Greenbelt Rd, Suite 400, Greenbelt, Maryland 20770, USA}
\author{Edward H. Chen} \email{ehchen@mit.edu}
\affiliation{Department of Electrical Engineering and Computer Science, 
Massachusetts Institute of Technology (MIT), Cambridge, Massachusetts 02139, USA}
\author{Gabriel A. Durkin} \email{Gabriel.Durkin@qubit.org}
\affiliation{Quantum Artificial Intelligence Laboratory (QuAIL), 
NASA Ames Research Center,  Moffett Field, California 94035, USA}
\affiliation{SGT Inc., 7701 Greenbelt Rd, Suite 400, Greenbelt, Maryland 20770, USA}
\date{\today}
\begin{abstract}
Quantum instruments derived from composite systems allow greater
measurement precision than their classical counterparts due to
coherences maintained between N components; spins, atoms
or photons. Decoherence that plagues real-world devices can be particle loss, or thermal excitation and relaxation, or dephasing due to external noise sources (and also due to prior parameter uncertainty). All these
adversely affect precision estimation of time, phase or frequency. By a novel technique we uncover the uniquely optimal probe
states of the N `qubits' alongside new tight bounds on precision under
local and collective mechanisms of these noise types above. For large
quantum ensembles where numerical techniques fail, the problem
reduces by analogy to finding the ground state of a 1-D particle in a
potential well; the shape of the well is dictated by the type and
strength of decoherence. The formalism is applied to prototypical Mach-Zehnder and Ramsey interferometers to discover the ultimate performance of real-world instruments.
\end{abstract}

\maketitle

Decoherence in quantum systems is responsible for a transition to
classical behavior. It degrades the advantage offered by quantum correlations for metrology, specifically the task of parameter estimation. 

An overview of the field of quantum metrology is provided in 
refs.~\onlinecite{Giovanetti04,Giovanetti11} and fig.\ref{cartoon} outlines the prototypical quantum metrological process, illustrated in parallel for a Mach-Zehnder interferometer and atomic clock. Previously, some of us
derived optimal states and metrological bounds for phase estimation subject to particle loss\cite{knysh}.
Refs.~\onlinecite{RDD12,diffother} additionally derive some
bounds for types of individual decoherence following ideas introduced by Fujiwara and Imai \cite{fujiwara}.
Subsequent work \cite{Escher12} employs the same approach to tackle
collective dephasing. These results are variational in nature and
guarantee neither tightness of the bounds, nor provide any intuition
about those probe states attaining best precision.
Numerical work by Genoni et al. \cite{Paris11} considers collective
dephasing and obtains optimal continuous variable states, restricted to those possessing a Gaussian characteristic function.

We consider for the first time in quantum metrology a very general physical model of decoherence that
includes processes of dephasing, relaxation, and excitation. These three categories are investigated via collective or individual mechanisms; an ensemble of $N$ particles is coupled either to a single common bath, or each to its own. For interferometry we explore relevant  processes of photon loss combined with collective dephasing, as might occur due to thermal motion of optical components or intrinsic laser noise.

We will show via a new operator formalism, presented in the Methods section, that for a large number of qubits $N$ the problem of identifying the unique and \emph{previously unknown} optimal states in line with tight precision bounds is mapped onto that of finding the ground state of a quantum-mechanical particle in a 1-D potential. Results are presented in table II.  

\begingroup
\squeezetable
\begin{table}
\begin{ruledtabular}
 \setlength{\tabcolsep}{3pt} 
    \renewcommand{\arraystretch}{2.5}
\begin{tabular}{l|c|c}

& Collective ($\dot\Gamma^\sigma\times$) & Individual ($\dot\gamma^\sigma\times$) \\
\hline
Dephasing: $\sigma \mapsto 0$ 
  & $S^z \rho S^z - \frac{1}{2}\{(S^z)^2,\rho\}$ 
  & $\sum_k s_k^z \rho s_k^z - \frac{N}{4} \rho$ \\
Relaxation: $\sigma \mapsto -$ 
  & $ S^- \rho S^+ - \tfrac{1}{2}\{S^+S^-,\rho\} $
  & $\sum_k \bigl( s_k^- \rho s_k^+ - \tfrac{1}{2}\{s_k^+s_k^-,\rho\} \bigr)$\\
Excitation: $\sigma \mapsto  +$ 
  & $ S^+ \rho S^- - \tfrac{1}{2}\{S^-S^+,\rho\} $
  & $\sum_k \bigl( s_k^+ \rho s_k^- - \tfrac{1}{2}\{s_k^-s_k^+,\rho\} \bigr)$ \\

\end{tabular}
\end{ruledtabular}
\caption{\label{tabnoise}
\textbf{Lindblad operators for various types of noise}, $\mathcal{L}_\sigma [\rho]$
are summarized in this table. Entires are multiplied by
appropriate rates of either collective ($\dot\Gamma^0$,
$\dot\Gamma^-$, $\dot\Gamma^+$) or individual
($\dot\gamma^0$, $\dot\gamma^-$, $\dot\gamma^+$)
dephasing, relaxation or excitation, respectively. Overdots denote
derivatives $\partial/\partial\theta$ with respect to `time-like' phase variable $\theta$.
}
\end{table}
\endgroup

We improve known precision bounds for individual relaxation by a factor of
two. It is also revealed that for collective dephasing the `quantum'
component of error (in excess of classical diffusion) turns out to be 
10 times ($\pi^2$) larger than suggested by previous reports \cite{Escher12}. 

Ultimately, all bounds we derive are achievable \emph{as they are constructive}; obtained via the discovery of a probe state that is uniquely optimal in the large ensemble limit $N \gg 1$. The asymptotic tightness of our bounds then follows naturally. The general form of this probe may be fairly exotic, but despite this, in many typical, combined noise regimes the optimal probes approach a Gaussian profile (albeit with mean and variance set by the `flavor' of decoherence). This is  pertinent when it comes to the challenge of generating these probes in the laboratory. For mixed noise types the predicted optimal states can be apparent for modest ensembles of $N \lesssim 30$ particles, indicating the power of this analytic approach whether the available resources are few or many. See fig.\ref{compare} for a comparison with numerics.

Finally, by fixing the total resources $\nu N$, where $\nu$ is the
number of measurement trials and $N$ is the size of each multi-qubit `cluster' (subjected independently to the system dynamics), we uncover novel scaling laws in table II. For individual decoherence there exist entangled states with particular structure that is most robust to noise; it is always the largest qubit clusters with this structure that offer the best precision. Collective decoherence, by contrast, is eventually deleterious to \emph{all}
entangled states. Beyond a critical cluster size $N_c$ overall performance decreases. At the critical size, clusters of NOON\cite{NOON} states -- despite their recent reputation for fragility to decoherence\cite{NOON-crap} -- are resurrected as an optimal metrology resource.

\section*{Phase Estimation with Decoherence}
The precision of parameter estimation is controlled by three factors: (i) the input state or probe $| \psi \rangle$ subjected to (ii) dynamical evolution which imprints the parameter $\theta$ to be estimated, and (iii) the specific measurement choice that reveals details of the evolved state and therefore the parameter, see fig.1. A fourth, often overlooked aspect is the need to combine many repeated measurement outcomes into a minimum-variance, unbiased estimator. Classical Fisher Information
(CFI) gives a measure of precision for such an efficient estimator without choosing the estimator explicitly.  Quantum Fisher Information (QFI)  then represents CFI optimized over all possible
measurements \cite{Braunstein94,ParisQmTech}. The only remaining step in deriving the ultimate
bounds (for fixed system dynamics) is optimization over the probe state using QFI as a precision metric. This is our task in the present work.

\begin{figure}
\includegraphics[width=3.5in]{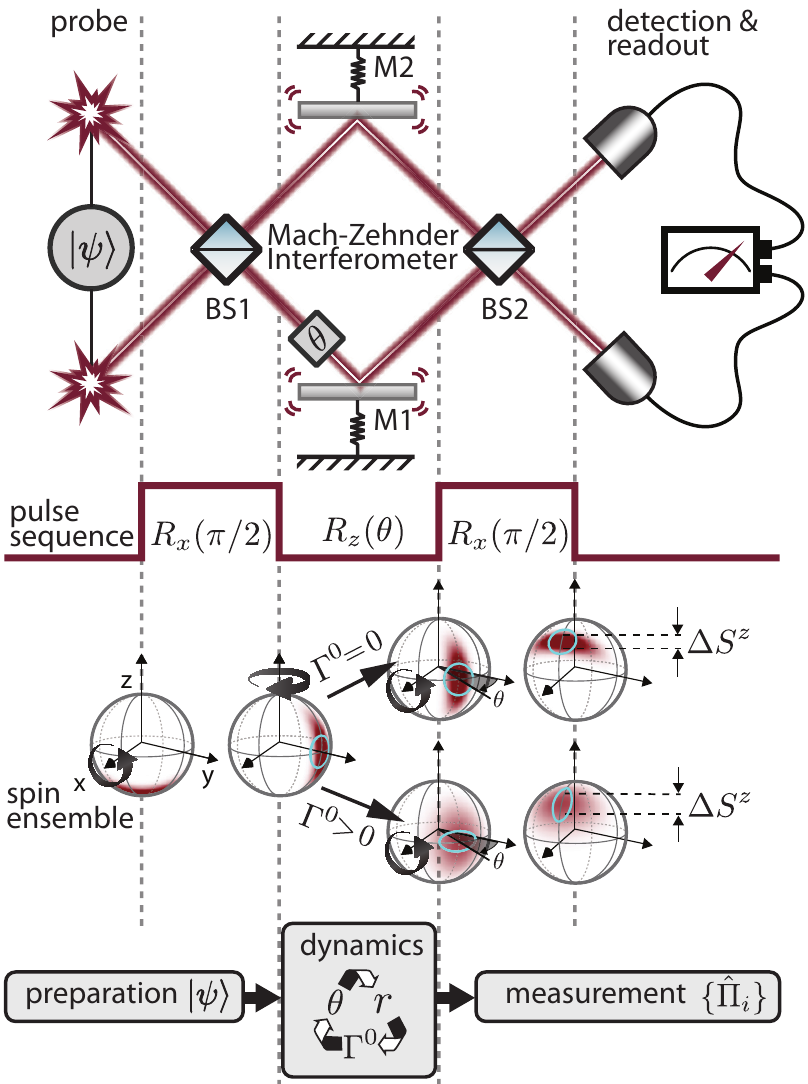}
\caption{\label{cartoon}
\textbf{Quantum phase estimation} is illustrated for optical (Mach-Zehnder) and 
atomic (Ramsey) interferometry. Process diagram beneath indicates three physical steps of quantum metrology: probe state preparation, dynamical evolution, and measurement. Not shown is the final statistical analysis of measurement results to produce a parameter estimation. The Schwinger isomorphism maps accumulation of relative phase $\theta$ 
due to interferometer path difference to temporal evolution of  $N$ two-level atoms or spins
for time $\theta$, i.e. free rotation $R_z(\theta)$. Actions of beamsplitters BS1, BS2
are equivalent to $\pi/2$-pulses producing rotations about $x$-axis.
Red density plots on spin spheres depict probability (Husimi\cite{Nori}) distributions for $N=60$
phase-squeezed state of eqn.\eqref{opt}. For comparison, blue circles indicate the 
boundary of probability distribution for the spin-coherent 'pointer' state, for which all $60$ spins are aligned. Increased projection
measurement error $\Delta S^z$ due to dephasing $\Gamma^0 =0.05$ is indicated in the lower branch ($S^z$ measurement may be performed by photon-counting in both Mach-Zehnder arms after BS2, or  $S^x$ measurement between BS1 and BS2.). Dephasing may occur at interferometer mirrors M1, M2
by thermal  motion or radiation pressure. Other sources of noise include photon loss $r$.}
\end{figure}

Dynamical evolution of an ensemble of $N$ qubits can be described by a master
equation\cite{mastereqn}:
\begin{equation}
  \frac{\dd \rho}{\dd \theta} = -\ii [S^z, \rho] + 
  \underbrace{\textstyle\sum_\sigma \mathcal{L_\sigma}[\rho]}_\text{noise terms},
  \label{ME}
\end{equation}
where unitary shift operator $S^z = \sum_{k=1}^{N} s_k^z$ conserves total
spin $S \leqslant N/2$: the eigen-equation is $S^z | S,m \rangle = m | S,m \rangle$.
The largest, $S=N/2$ space is completely symmetric 
under particle exchange; for bosons (e.g. photons) this is the only possibility.
This spin formalism is applicable to ensemble of atoms in a double-well potential 
\cite{BECMZ}, optical interferometers \cite{Yurke}, or Ramsey interferometry used 
in atomic clocks \cite{clocks} (see fig.~\ref{cartoon}).

Absent noise, minimum variance of phase estimation is inversely proportional to
the variance of phase shift operator $S^z$:
\begin{equation}
  \nu \var \theta_\mathrm{est} = 1/F, \qquad F=4 \var S^z,
\end{equation}
where $\nu$ is the number of independent measurements. Optimal probe states
inhabit the subspace of largest  $S=N/2$. They have 
$\langle  S, \pm S | \psi\rangle = \psi_{\pm S}=1/\sqrt{2}$ as the only non-zero amplitudes: a celebrated `GHZ' \cite{GHZ} 
aka `NOON' \cite{NOON} probe state. Also note that, as $\var \theta_\mathrm{est} =
1/(\nu N^2)$, it becomes advantageous to utilize the largest ensembles possible, trading 
off the number of measurements for larger $N$ when the total qubit resources $\nu N$ 
are constrained.

\begin{figure*}
\includegraphics[width=5.2in]{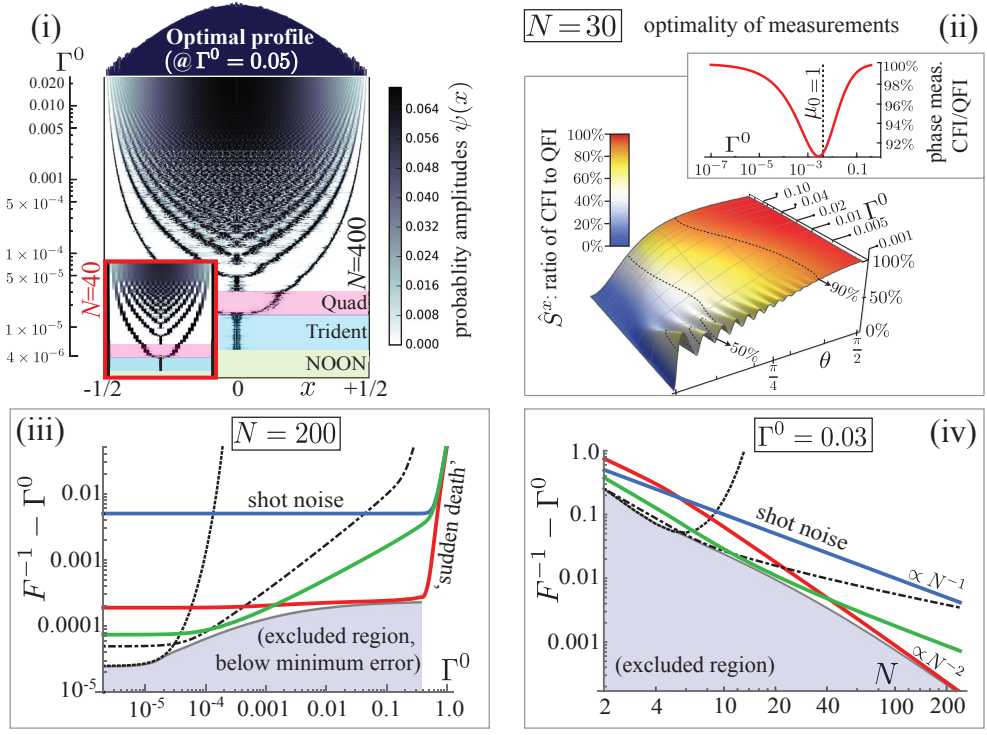}
\caption{\label{quad}
\textbf{Collective Dephasing has an associated optimal state} that evolves via a characteristic `Menorah' structure in (i) for $N = 400$ (inset for $N=40$) as dephasing increases (in the vertical direction). Each horizontal row of pixels  $(x = m/N)$ correspond to the array of optimal state components $\psi_m$ for $m \in [-N/2, +N/2]$  found by numerical search; each component's amplitude indicated by pixel brightness. As dephasing increases a third component appears at $m=0$ in addition to the  two extremal ($m = \pm N/2$) NOON components to make a `Trident' optimal state. That central component in turn bifurcates  symmetrically at the next critical point to make a `Quad' state; those two new components diverge towards the boundary. By such a series of accelerating bifurcations the optimal state eventually form a continuous Cosine profile, eq. \eqref{opt}, slightly ragged but already apparent at $\Gamma^0 = 0.05$ (upper barchart). Figure (ii) gives optimality (ratio of CFI to QFI) of phase measurements (2D line plot) and $S^x$ or photon counting (3D surface) for the state of eq.\eqref{opt} and $N=30$. Phase measurements exhibit a dip, they are optimal for both large and small mass  $\mu_0 = \Gamma^0 N^2$, but not for intermediate regime $\mu_0 \approx 1$. For $S^x$ measurements efficiency depends on the phase neighborhood as well as dephasing $\Gamma^0$. Biasing the interferometer phases $\theta \approx \pi/2$ allows such measurements to be close to optimal for all $\Gamma^0$.   Plots (iii) and (iv) show quantum measurement error, from eq.\eqref{Fadd}, for different families of states.  The asymptotically optimal Cosine state (red curve) is compared with phase states \cite{vourdas} (green) with $\psi_m = 1 / \sqrt{N+1}$  and Holland-Burnett states
\cite{Holland93} (black chained line) with $\psi_m =  d^{S}_{m,0}(\pi/2)$ (note.~\onlinecite{wigner}). NOON states (black dotted) gain advantage for very small dephasing $\Gamma^0 \lesssim 10^{-5}$. A spin-coherent state
[blue curve, $\psi_m = d_{m,S}^{S}(\pi/2)$] is operating
at the shot-noise limit exactly, $1/N$. Note the `sudden death' of precision as $\Gamma^0 \gtrsim 1$, when the
density matrix is nearly diagonal and symmetric under shift by $S^z$. The phase-squeezed state \eqref{opt} is already dominant for moderately small $N \gtrsim 40$ in (iv) and across three orders of magnitude of $\Gamma^0$ in (iii). Lowest possible quantum error is found numerically; indicated by the upper boundary of the `excluded' region.}
\end{figure*}

The situation certainly becomes more complex when any decoherence effects are included. 
An initially pure state decays into a mixture due to coupling to the environment,
which can be modeled by adding Lindblad terms within eq.\eqref{ME}  (summarized in table~\ref{tabnoise} ).
We are careful to distinguish individual and collective decoherence. The former would 
be appropriate when each qubit is coupled to its own bath, while the latter results in 
much stronger noise when all qubits share common bath and have the same
coupling constant. A consideration of weak decoherence involves the effect of phase diffusion alone;
 characteristic ensemble dephasing time $T_2^\ast \propto1/ \dot\Gamma^0$ is much less than the characteristic relaxation time $T_1 \propto  1/ \dot\Gamma^-$.  Many realistic scenarios will involve stronger relaxation, and we shall consider this too. Excitation will be included for completeness, playing a role at finite temperatures.
It should be emphasized that this is a fairly exhaustive list of
decoherence processes. For instance, depolarization considered in 
refs.~\onlinecite{fujiwara,RDD12,DepolChan} is not a distinct process but can be
described as a combination of dephasing, relaxation and excitation
e.g. with $\gamma^0/2 = \gamma^- = \gamma^+$.

In the present work we eschew \emph{ad hoc} precision bounds in favor of
analysis of dynamics in the limit $N \gg 1$ where the major gains of
quantum metrology may be realized, if phase error is shown to scale with some negative power of $N$. 
The optimal state
may be approximated by a wavefunction 
$\langle N/2, m | \psi \rangle \mapsto \psi(x)/\sqrt{N}$ [where $x=m/N$ is
taken to be continuous] corresponding to the ground state of a second order differential equation akin to the time-independent Schr\"odinger equation (see Methods section). Following the analogy, the ground state `energy' eigenvalue $\lambda_\text{min}$
determines the minimum attainable phase variance:
\begin{align} 
  -\psi_\text{opt}'' + \mu(x) \psi_\text{opt} &= \lambda_\text{min} \psi_\text{opt} \: , \nonumber \\
  \quad \var\theta_\text{est} & \geq \lambda_\text{min} / (\nu N^2) \: .
  \label{schrodinger}
\end{align}
The form of the potential well $\mu(x)$ depends on the flavor and strength of decoherence, see results of table II. It should also
include infinite walls $\mu(x)=+\infty$ for $|x|>1/2$ to ensure that the
wavefunction vanishes outside the interval: $-N/2 \leq m \leq N/2$.

\begin{figure}
\includegraphics[width=2.8in]{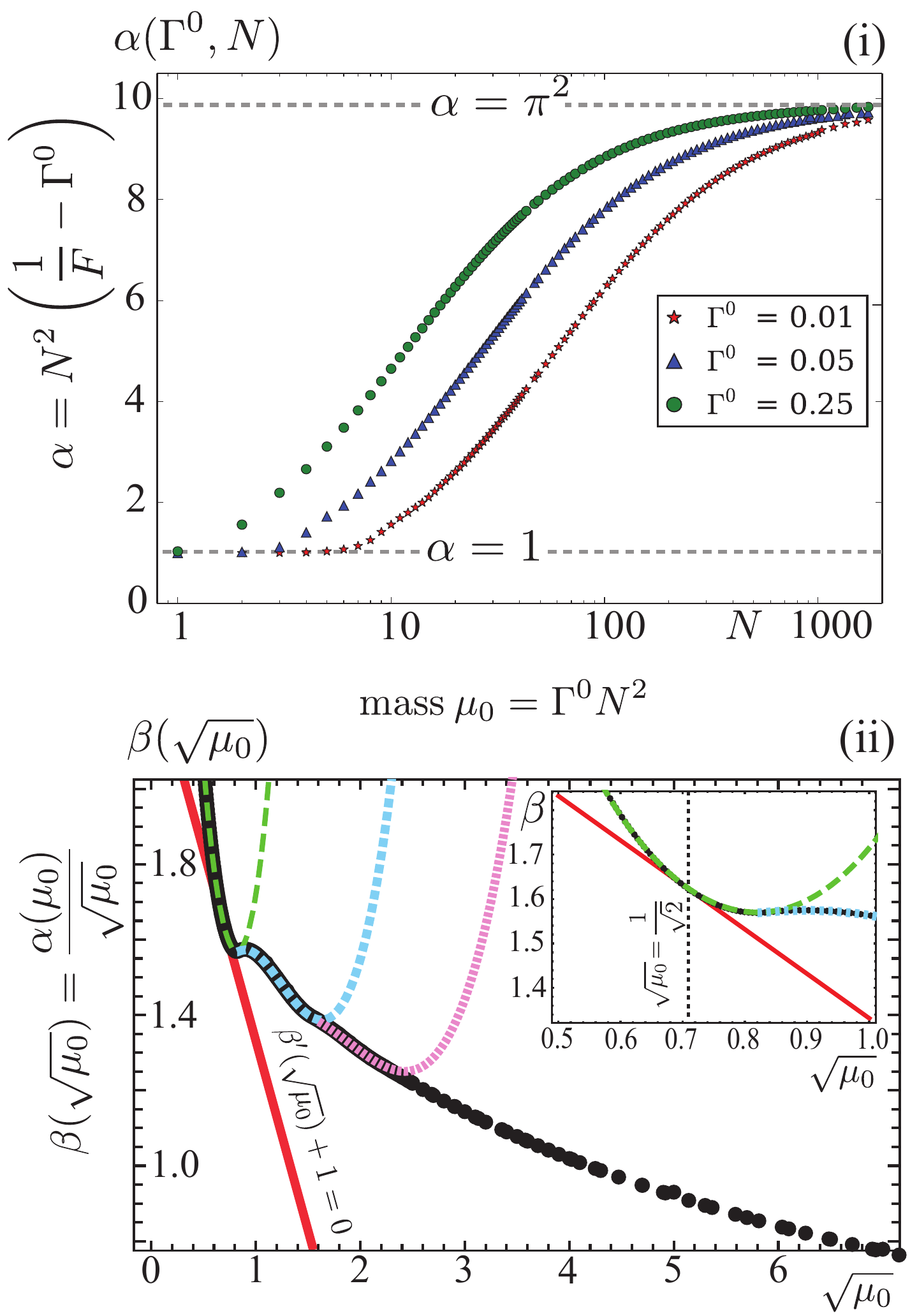}
\caption{\label{alphabeta}
\textbf{Numerical Results} for collective dephasing. In (i) individual plot points correspond to numerical searches for the minimum coefficient $\alpha$ of the measurement error contribution $\alpha / N^2$ to mean-squared phase error. Red stars, blue triangle and green circles are for dephasing  $\Gamma^0 \mapsto \{0.01,0.05,0.25\}$, respectively. Coefficient $\alpha$ interpolates from unity to $\pi^2$ [from the bounds of eq.\eqref{bounds}] as universal `mass' parameter $\mu_0 = \Gamma^0 N^2$ increases. In graph (ii) previous numerical data collapses onto a single curve-- indicating dependency on a single combined parameter, $\mu_0 = \Gamma^0 N^2$ (black points). The  $\alpha$ value for optimal clustering is located where function $\beta(\sqrt{\mu_0})$ has gradient $= -1$, corresponding to $\mu_0 = 1/2$ (inset, vertical dotted line). The $\beta$ function is plotted for NOON (green), `Trident' (light-blue) and `Quad' states (pink), indicating that optimal clusters will be NOON states. (See discussion in the Methods.) The minima of these analytic curves correspond to the bifurcation points in the Menorah of fig.\ref{quad}(i).}
\end{figure}

We have corroborated our results with numerical study, presented in figures: \ref{quad}, \ref{alphabeta}, \ref{compare}. As seen in fig.\ref{quad}(i) and fig.\ref{compare}, the characteristic asymptotic behavior will often `set in' for relatively small particle number $10<N <100$, allaying concerns that these results are of interest only  in the limit of very large ensembles. To contrast, if we had na\"ively chosen to optimize precision bounds appearing elsewhere in the literature\cite{diffother,Escher12}, we would recover NOON-like  `optimal' probes having only a few discrete
components (states that actually perform poorly in noisy conditions), rather than the necessary continuous profile.

\section*{Collective Dephasing}

In the following sections we describe various noise combinations in
more detail, but let's begin with the simplest scenario, that of
`pure' collective dephasing. 
It is the dominant noise in atom and spin ensembles, where energy and particle number are conserved \cite{dephasing10}. It occurs in light beams due to laser noise, optical path length fluctuations \cite{Roman06} and radiation 
pressure at the surface of mirrors \cite{Caves80,Escher12}. It plays a role in 
optic-fiber interferometric sensors \cite{fibershapesensor} where thermal 
perturbations and mechanical strains can lead to measurable diffusion in both
interferometric phase and polarisation of light. Collective dephasing
is the prevalent noise for ions confined to traps \cite{Blatt14Qubits}. 

Dephasing presents an exponential suppression of the off-diagonal
elements of the density matrix. Equivalently it can be viewed as random fluctuations of the phase $\theta$ itself with amplitude $\sqrt{\Gamma^0}$:
\begin{equation}\label{uncert}
\rho = \int_{-\infty}^{\infty}  | \psi (\theta) \rangle \langle \psi (\theta)  | \frac{\ee^{-(\theta-\bar\theta)^2/ 2  \Gamma^0}}{\sqrt{2 \pi \Gamma^0}}     \dd \theta.
\end{equation}
The density matrix is a mixture of pure states $|\psi\rangle$, each evolved by a different phase $\theta$, but the ensemble is Gaussian-distributed with mean
$\bar\theta$ and variance $\Gamma^0$. The statistical ensemble may form via independent measurements on many identical pure states, or may result simply from coupling to the environment. Overall, environmental dephasing is indistinguishable from an acknowledgment of initial phase uncertainty, even when noise is absent, indicating the ubiquity and central role of collective dephasing in the estimation process.

The quantum contribution to phase error under dephasing is a subtle, second-order $(\propto 1/N^2)$ correction to the classical phase noise. Attempts to find optimal states by WKB approximation were defeated and the solution required careful application of a novel operator approach, presented in Methods. (We identify the dephasing channel with an imaginary-time Feynman propagator for a free particle.) Further
details of the calculation can be found in Appendix A.  We find that the optimal state and minimum variance are
described by eqn.~\eqref{schrodinger} with a `flat' potential
$\mu(x) = \mu_0 = N^2 \Gamma^0$ between the infinite walls at $x = \pm 1/2$. Let us call this large parameter `mass'. A semiclassical asymptotic expansion of QFI maybe carried out in powers of $1/\mu_0$; truncating the series after the first few non-trivial terms is the basis of many of our results. For large `mass' $\mu_0 \gg 1$, the optimal state is given by the ground state of a `particle in a box', a
\emph{phase-squeezed} Cosine function spanning half a period. Expressed
in terms of discrete amplitudes,
\begin{equation}
  | \psi_\text{opt} \rangle = \sqrt{\frac{2}{N+1}} 
  \sum_{m=-N/2}^{N/2} \cos \frac{\pi m}{N+1} \bigl| \tfrac{N}{2}, m \bigr>.
  \label{opt}
\end{equation}
(This state was originally proposed in ref.~\onlinecite{PeggSummy} as a candidate for precision in the absence of noise.)
Phase error approaches $\nu N^2 \var \theta_\text{est} = \mu_0+\pi^2$ 
in the limit of large $N$. Additionally, we constrain ourselves to a
scenario where $\Gamma^0 \ll 1$; large values severely suppress
phase-carrying off-diagonal elements of the density matrix.
We shall see also that entangled ensembles offer no advantages when noise
exceeds certain threshold $\Gamma^0_c \lesssim 1$.

Convergence of the optimal probe from a discrete NOON state to the continuous phase-squeezed profile above is observed numerically in fig.\ref{quad} (i), progressing by a sequence of accelerating bifurcations as dephasing increases. Fig.~\ref{quad}.(iii) and (iv) give quantitative comparison with families of states proposed in the literature.

A way to generate the optimal phase-squeezed state was presented by
Combes and Wiseman \cite{Josh}. The state is generated by the action of a 
counter-twisting spin-squeezing Hamiltonian $S^y S^z + S^z S^y$ 
(ref.~\onlinecite{spin-sq}) on an unentangled spin-coherent state.
An exotic physical mechanism was given by Andre and Lukin  involving interaction 
between atoms via polaritons \cite{LukinSpin}. Yet another approach utilizes the 
dynamics of a two-mode Bose-Einstein condensate for optimal phase
squeezing \cite{BECoptimal}.

\subsection*{Combining Errors from Classical Noise and Measurement}

An optimal state \eqref{opt} can be found by minimizing the functional
for the reciprocal of QFI (see the Methods),
\begin{equation}
  \nu \var \theta_\text{est} = \frac{1}{F} 
  \approx \Gamma^0 + \frac{1}{N^2} \int \psi'^2(x) \dd x,
  \label{Fadd}
\end{equation}
valid in the `large-mass' regime $\mu_0 \gg 1$. Notice that to this lowest order the phase uncertainty is a sum of errors added in quadrature; the amplitude $\Gamma^0$ is the variance of a classical random phase noise or equivalently that of a (Gaussian) prior uncertainty, recall eqn.\eqref{uncert}. The second term is the quantum measurement uncertainty, as follows: 

In this large mass regime canonical phase measurements, i.e. projections onto the (over-complete) basis of phase states $| \theta \rangle = \tfrac{1}{\sqrt{N+1}} \sum_m \ee^{-\ii m \theta} \bigl| \tfrac{N}{2},m \bigr>$, are optimal \cite{supplement}; they produce a classical Fisher information equal to the QFI. Because phase states are non-orthogonal they produce a finite-width distribution of measurement error; the phase distribution of the probe must be \emph{convolved} with the classical phase uncertainty of width $\sqrt{\Gamma^0}$  from eq.\eqref{uncert} when dephasing (or prior uncertainty) is present. The convolution of two Gaussians produces another with variance equal to the sum of the two components variances. Hence  Gaussian profile probe states (with a Gaussian-distributed phase distribution) produce  \eqref{Fadd} exactly. Such is the spin-coherent state obtained in fig.\ref{cartoon} between the beam-splitters when a photon-number state enters just one port of the first beamsplitter. The quantum measurement mean-squared error for this state is $1/N$ [see fig.\ref{quad} (iv)]; poor `shot-noise' scaling can be
explained by vacuum fluctutations leaking into the other `dark' port \cite{Caves81}. 

For generic probes, \eqref{Fadd} may still be applied when
Gaussian-distributed `classical' phase fluctuations are the dominant error
source. In this scenario, to maximize precision and QFI one must minimize the phase variance of the quantum measurement component in \eqref{Fadd}, since  $\Gamma^0$ is fixed. The state to minimize phase measurement error (in the absence of noise \cite{PeggSummy}) is  \eqref{opt}, hence its appearance above (and its description as a `phase-squeezed' state). Optimality of this state is inevitable for increasing dephasing \emph{or} ensemble size $N$, as the relevant universal parameter is the product $\mu_0 = \Gamma^0 N^2$. 

In the opposite `small-mass' regime $\mu_0 \ll 1$, the GHZ/NOON state
achieves the best precision -- as in the noiseless scenario. The mean squared quantum
error component is $1/N^2$ or about 10 times smaller than that
achieved with the `minimum phase variance' (phase-squeezed) state \eqref{opt}.
It is a subtle point that this reduction is possible by a more efficient estimator than the sample mean, (that was only sufficient for $\mu \gg 1$ because the final phase distribution was approximately Gaussian). Interestingly, canonical phase measurements are optimal in both small and large mass regimes but not in the intermediate regime $\mu_0 \approx 1$; observe the dip of fig.\ref{quad}(ii).

We must add that such measurements
are not a unique choice: projections onto eigenstates of $S^x$ can also saturate these bounds in the
limits of either small or large mass $\mu_0$. Examine the surface plot of fig.\ref{quad} (ii) indicating performance of $S^x$ measurements as a function of both phase and dephasing. Because canonical phase measurements are much more difficult to implement,\cite{phase-meas} it is significant that $S^x$ measurements can be optimal-- they correspond to photon counting in interferometry or ensemble `population difference' measurement in an atom clock .

In the intermediate-mass regime, minimum error interpolates between two bounds
\begin{equation}
  \Gamma^0 + 1/N^2 \leqslant 1/F \leqslant
  \frac{\Gamma^0+\pi^2/N^2}{[1-\pi \tilde p(\pi)]^2} \: ;
  \label{bounds}
\end{equation}
optimal phase estimation requires coordination of a highly non-trivial probe state,
measurement, and estimator statistic. The first inequality can be understood as a Fisher
information inequality for the sum of two random variables:
$F_{x+y}^{-1} \geqslant F_x^{-1} + F_y^{-1}$ (where $x$ and $y$ are respectively the true interferometric phase and the random fluctuating phase), appearing previously in 
ref.~\onlinecite{Escher12}. The upper bound on $1/F$ above has been modified
via a denominator that takes phase periodicity into account \cite{UncertPhase}.
[Here $\tilde p(\theta)$ is the probability distrbution obtained by convolving
the distribution of random phase from eq.\eqref{uncert} and phase measurement error, given explicitly in Appendix A, eq.\eqref{pdth}].
The effect of this denominator can be ignored in the limit of small dephasing, but leads to error
increasing in proportion to $\exp(\Gamma^0)$ as soon as $\Gamma^0 \gtrsim 1$, a `sudden death' of precision. See fig.\ref{quad}(iii).

\subsection*{Clustering and Shot-Noise Scaling} \label{cluster}

For a fixed \emph{total} number of resources $\nu N$,
increasing $N$ is done at the expense of the number of independent
measurements $\nu$, increasing the variance due to classical phase diffusion limit
$\Gamma^0 / \nu$. The `quantum' component of error decreases owing
to  Heisenberg scaling $N^{-2}$. Optimal cluster size $N_c$ is
determined by a tradeoff between these contributions. The minimum
phase estimation variance for this optimal choice becomes
$c\sqrt{\Gamma}/(\nu N)$ in the limit of small $\Gamma^0$. 
The value of the prefactor can be bounded between $2$ and $2\pi$ by
performing independent optimizations of on both sides of double
inequality \eqref{bounds}. Obtaining the precise value requires the
knowledge of the functional dependence of $\alpha(\mu_0)$ of the
prefactor of quantum component of error $\alpha/N^2$ since optimal
cluster size $N_c$ necessarily corresponds to the intermediate mass
$\mu \sim 1$. We find (in fig.\ref{alphabeta}) that $c=\sqrt{2\ee}$, corresponding
to $N_c = 1/\sqrt{2\Gamma^0}$ (`mass' $\mu_0=1/2$). A NOON state is
still the optimal choice of probe at this mass value.
The unexpected `resurrection' of NOON state as the optimal probe needs more
explanation. While it is true that NOON states are extremely fragile
in the presence of decoherence,\cite{NOON-crap} collective dephasing degrades the
performance of \emph{all} states with large $N$. For decoherence of
individual character, precision always improves for increasing $N$ but
requires employing exotic probes that are more robust to noise than NOON states. With collective dephasing,
the only strategy to cope with noise is to refrain from using large
ensembles $N \gtrsim N_c \sim 1/\sqrt{\Gamma^0}$. And for large
dephasing, entangled states may offer no benefits at all. By comparing
the exact QFI expressions for a spin $S=1$ system and two unentangled $S=1/2$
particles, $2 \exp{(-\Gamma^0)}$, there is a critical dephasing $\Gamma^0_c
\approx 0.251$ beyond which sending $N$ particles one at a time is
better than using entangled clusters (even bipartite). By similar comparisons, tripartite
($S=3/2$) and 4-part ($S=2$) clusters become superior for
$\Gamma^0<0.081$ and $\Gamma^0<0.041$, respectively.

\begin{figure}
\includegraphics[width=3.2in]{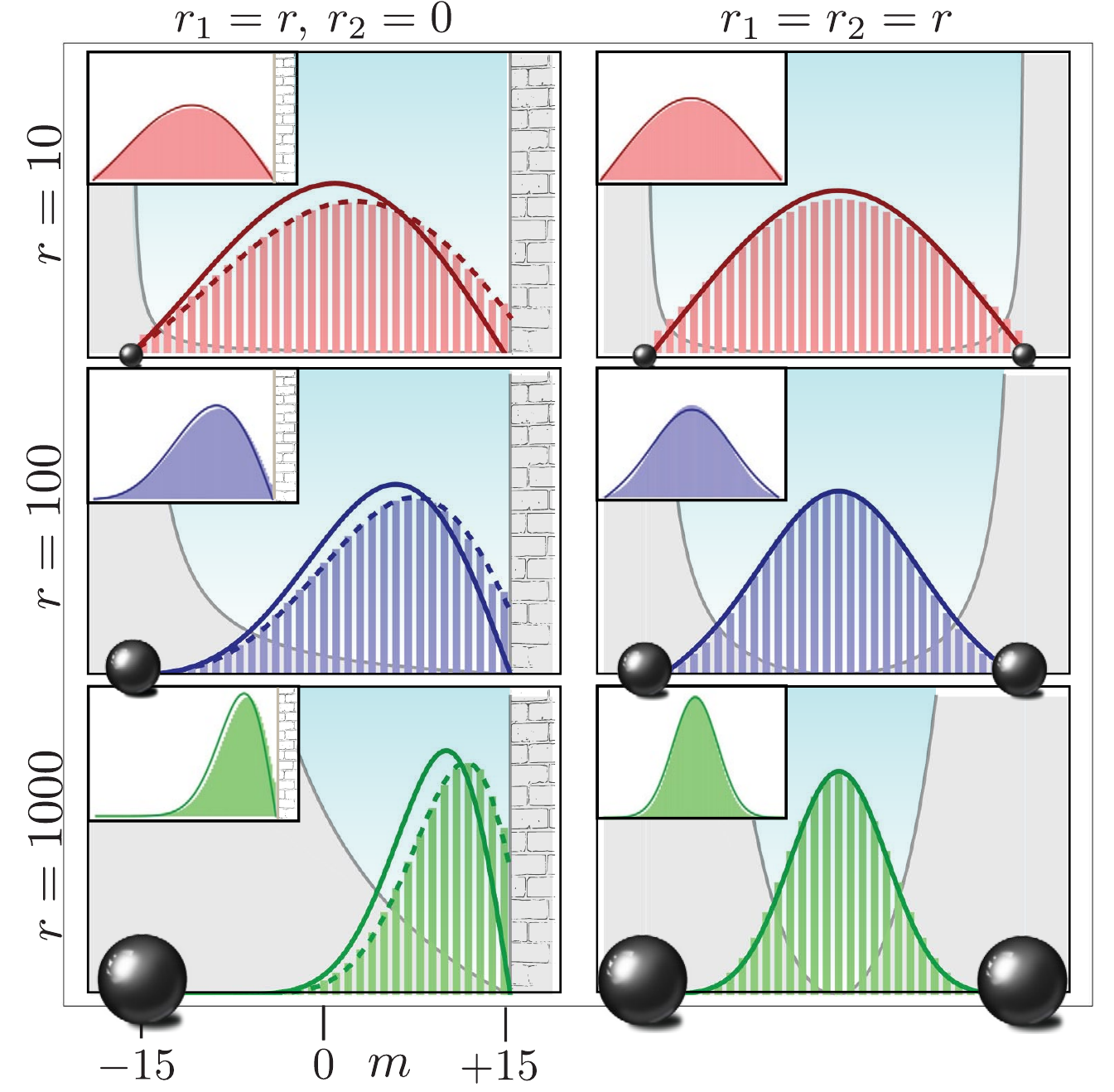}
\caption{\label{compare}
\textbf{Dephasing and particle loss -- Comparing analytics and numerics} in an interferometer for $N=30$  (inset $N=100$) and $\Gamma^0 = 0.25$; optimal state components are shown by solid bars. The left column is for losses in the sample arm only, $r_2 = 0$, and the right column for symmetric loss, $r_1 = r_2$. Increasing loss parameter $r \mapsto 10, 100, 1000$  for $N=30$ corresponds to reduced transmittivities $=75\%, 23\%, 3\%$.
Analytical solutions $\psi$ are superimposed as continuous line plots --  the ground state of a particle in a 1-D potential well (gray shaded region). This well is created by two repulsive Coulomb point sources, of `charge' $\propto r$. If $r_2 =0$  the second Coulomb source is replaced by an infinite wall. The  ground-state in that case is given by a Whittaker function with 
imaginary arguments $\psi(y,\eta) \propto y \ee^{-\ii y/2} 
{}_1M_1(1-\ii \eta, 2; \ii y)$ [note \onlinecite{fn4}] with $y=\lambda (1+2 m/N)$
and $\eta=r/8\lambda$. Boundary conditions to uniquely specify $\psi$ in $N \sim  \infty$ limit requires setting it to zero at Coulomb sources and the base of the wall. For  finite $N$ we can treat the Whittaker function as a `variational ansatz' rather than an analytic result, and optimize over the single parameter $\lambda$ (dashed curves), relaxing the `wall' boundary condition. For larger $N$ the node quickly converges on the point $m/N = x=1/2$ and the exact solution is recovered --  compare the inset $N=100$ data. In the large $N$ limit there can be no non-zero amplitude at the boundary -- Any discontinuity in $\psi(x)$ at $x=1/2$ causes an anomalous spike in kinetic energy $\propto \psi'(x)^2$, raising the solution out of the ground state of eq.\eqref{schrodinger}. In the symmetric-loss case the optimal state converges to a Gaussian profile of width $= (2 r^{1/4})^{-1}$ when loss is the dominant decoherence, $r \gg \mu_0$.}
\end{figure}

\begingroup
\squeezetable
\begin{table*}
  \begin{ruledtabular}
   \setlength{\tabcolsep}{1.5pt} 
    \renewcommand{\arraystretch}{3}
    \begin{tabular}{ c||c|c|c|c} 
%      \hhline{=||=|=|=|=}
      & \multicolumn{2}{c|}{Collective Decoherence}  & Individual Decoherence & Hybrid (Interferometry) \\
%      \hhline{=||=|=|=|=}
      \hline
      \hline
     Decoherence Process & Dephasing Only ($\Gamma^0$) & General Case
     ($\Gamma^0, \Gamma^{\pm}$)& General Case ($\gamma^0, \gamma^{\pm}$) & Collective Dephasing, Loss ($\Gamma^{0},\gamma_{1}, \gamma_2$)\\ 
      \hline
      Universal Parameters & 
     $\mu_0 = N^2 \Gamma^{0}$ 
      & $\mu_0$, and $\mu_1 = N^2 (\Gamma^+ +\Gamma^-)$& $r = N \left[
        \ee^{\gamma^0 + \gamma^- + \gamma^+}  -1 \right] $ & $\mu_0 \: $,  $r_{1,2}  = N \left[ \ee ^{\gamma_{1,2}}  -1 \right] \equiv
      4 N \epsilon_{1,2}^2$ \\ 
      \hline
      Potential $\mu(x)$ for $N \gg1$& 
      $\begin{cases} \mu_0 & -\tfrac{1}{2} < x \; <  \tfrac{1}{2} \\
        \infty & \; \;\; \; \; \; \; \;\;   |x| \geqslant \tfrac{1}{2} \end{cases}$ &
%      $\mu_0 \:  \forall \: |x|<\frac{1}{2}$  (box)& 
      $\mu_0 + \mu_1
      \dfrac{x^2}{1/4 - x^2} $ & 
      $\dfrac{r}{1-4 x^2}$ & $\mu_0 + \dfrac{1}{4} \left( \dfrac{r_1}{1/2 +x} +  \dfrac{r_2}{1/2  -x}  \right)$ \\ 
      (shape)&(box)&&(Coulomb-symmetric)& (Coulomb) \\
      \hline
      $\psi_{\text{opt}}$ profile: \{center, width, shape\} & \{$0,  \approx 0.21$, Cosine\}& \{$0 \: , \: (2 \mu_1^{1/4})^{-1}$, Gaussian\} &  \{$0\: , \: (2r^{1/4})^{-1}$, Gaussian\} &  \{$\frac{\sqrt{r_1}-\sqrt{r_2}}{2(\sqrt{r_1}+\sqrt{r_2})} \: , \: \frac{(r_1 r_2)^{1/8}}{\sqrt{r_1}+ \sqrt{r_2}}$, Gaussian*\}  \\ 
      Gives m.s.e. lower bound  & $   \Gamma^0 + \dfrac{\pi^2}{N^2}$ & $  \Gamma^0 + \dfrac{\sqrt{\Gamma^- +\Gamma^+}}{N}$  & $\dfrac{\exp (\gamma^0 +\gamma^{-}+\gamma^{+})-1}{N}$& $ \Gamma^0 + \dfrac{\left(\epsilon_1+\epsilon_2 \right)^2}{4N}$ \\ 
      (for conditions)& ($\mu_0 \gg 1$) & ($ \mu_0 \gg
      \sqrt{1+\frac{\mu_1}{\mu_0}}$ and $N \ll 1/\Gamma^{\pm}$) & ($r \gg 1$) & ($r_{1,2} \gg 1$) \\  
      \hline
      Optimal clustering bound ($N\!=\!N_c$) & 
      \multicolumn{2}{c|}{$\dfrac{\sqrt{2\ee\Gamma^0}}{ N}$} &  
%      $\alpha_1 \sqrt{\Gamma^0+\Gamma^-+\Gamma^+ }/  N$ & 
      n/a & $ \frac{(\epsilon_1+\epsilon_2)^2}{N} \! +
      \!  \begin{cases}[2.5]
        \frac{4}{N} \bigl( \tfrac{|a_1|}{4} \bigr)^{\frac{4}{3}} 
        \epsilon_1
        (\Gamma^0)^{\frac{1}{4}}\\
        \tfrac{3}{2^{2/3}} 
        \frac{(\epsilon_1 + \epsilon_2)^{\frac{4}{3}} (\Gamma^0)^{\frac{1}{3}}}{N(\epsilon_1\epsilon_2)^{1/3}}
        
      \end{cases}$ \\[12pt] 
    \end{tabular}
  \end{ruledtabular}
\caption{\label{resultstable}
\textbf{Quantum Precision Limits}: Summary of main results derived in this paper. Widths of optimal states are rescaled for variable $x = m/N \in [-1/2,1/2]$; multiply by $N$ for original scaling. The optimal state for interferometry is marked Gaussian* as it has a Gaussian profile when $r_{1,2} \gg 1$; however when $r_2 = 0$ and $\mu_0 \gg 1$ the profile approximates an Airy-type profile\cite{knysh} at large $r_1$. The last two rows show mean-squared error lower bounds on $\nu \theta_{\text{est}}$ for $\nu$ independent measurements; the final row has been optimized for a large fixed \emph{total} number of resources $\nu N$. Note that for individual decoherence there is no optimal clustering $N_c$ for a fixed  $\nu N$  as performance always increases monotonically with $N$. In interferometry, the Airy/Gaussian shape of the optimal profile in the large loss limit gives a modified phase error for fixed $\nu N$. This proceeds via next-to-leading order contributions to the potential $\mu(x)$ --  resulting phase error is shown in the final table element for both single mode, $\epsilon_2 = 0$ (upper) and two-mode loss cases, $\epsilon_2 > 0$ (lower). Constant $a_1 \approx -2.33$ is the first zero of the Airy function.}
\end{table*}
\endgroup

\section*{Interplay of Dephasing, Relaxation and Excitation}

For most combinations of
$\Gamma^\sigma$ for $\sigma \mapsto \{0, +, -\}$, there
will exist a wide range of magnitudes of $N$ where
the optimal state is determined by solving Schr\"odinger-type equation
\eqref{schrodinger} in a new, curved `potential', (see the Methods). Relaxation and
excitation processes give rise to an additional contribution 
$N^2(\Gamma^- + \Gamma^+)x^2/(1/4-x^2)$ to the potential (see
table~\ref{resultstable}).
As $\Gamma^\pm$ increases, the optimal state interpolates between
the Cosine state for pure dephasing and a  state centered at
the origin with a Gaussian profile of width 
$1/\bigl[ 2(\Gamma^- + \Gamma^+)^{1/4} \sqrt{N} \bigl]$.
The quantum component of error is $\sqrt{\Gamma^- + \Gamma^+}/N$,
i.e. it exhibits shot-noise scaling. The next order term is  of order $1/N^2$, due to anharmonic
corrections  to the potential; it plays a role in determining the
optimal cluster size $N_c$.

This result holds only for $N \lesssim 1/\Gamma^\mp$. For larger ensembles the
dynamics predict a `\emph{super}-sudden death' of precision. Error
increases as $\exp[N(\Gamma^- - \Gamma^+)/2]$, an effect related to
Dicke's superradiance\cite{Dicke}: Coherent emission of quanta into a shared bath
results in much stronger dissipation.

Addressing the issue of optimal clustering, our findings are unchanged
from the case of collective dephasing. NOON states are robust with
respect to relaxation and excitation, except when $N \gtrsim 1/\Gamma^\pm$ 
where \emph{all} states perform poorly. On the other hand, degradation
of performance due to dephasing will occur as $N \gtrsim 1/\sqrt{\Gamma^0}$, prior to `super-sudden death' of precision in
the regime where all decoherence processes have comparable strength.
Minimum phase estimation variance is primarily determined by the
strength of dephasing with only a small correction due to other
processes, see table \ref{resultstable}.

\subsection*{Individual Decoherence}
Let's now consider a model where each qubit is coupled to its own bath,\cite{Huelga1997,local-decoh} rather than a single common bath.

We are primarily interested in a regime where the strength of
decoherence remains constant while the number of qubits grows large.
A somewhat surprising result (see the Methods) is that the ultimate precision and the
optimal state depend on the total strength of decoherence
$\gamma = \gamma^0 + \gamma^- + \gamma^+$ but not on 
fractions attributed to particular processes: dephasing, excitation,
and relaxation. The form of potential $\mu(x)$ is given in table~\ref{resultstable}.
Its minimum determines an asymptotically tight upper bound:
\begin{equation}
  \var \theta_\text{est} \geqslant \frac{\ee^\gamma - 1}{\nu N}.
  \label{indbound}
\end{equation}
 Approximating the potential by a parabola we find the optimal state
and the leading order correction 
$2\sqrt{r}/(\nu N^2) = O\bigl(\tfrac{1}{\nu N^{3/2}}\bigr)$.
Bound \eqref{indbound} coincides with the bound obtained earlier
for pure dephasing \cite{diffother,RDD12} but is \emph{stronger} than all known
bounds for relaxation and depolarization (\emph{ibid}).

\section*{Application to Interferometry}
In a two-mode interferometer, collective dephasing takes place
concurrently with particle loss, in effect a hybridized individual/collective noise. Photons are easily lost in optical components but atoms may also be absorbed into the surrounding thermal cloud during the decay of a Bose-Einstein condensate \cite{BECloss}.

Lindblad operators $\dot \gamma_k \bigl(
  a_k^{} \rho a_k^\dag - \tfrac{1}{2} \{ a_k^\dag a_k^{},\rho \} \bigr)$,
with $k=1,2$ represents losses in sample and reference arms
respectively with rates $\dot \gamma_{1,2}$. (In fig.\ref{cartoon} these are the modes with mirrors $M1$ and $M2$ respectively.) Here $a_k$, $a_k^\dag$
are particle annihilation and creation operators.  Beamsplitters with transmission $\ee^{-\gamma_1}$ and 
$\ee^{-\gamma_2}$ couple particles (photons) into
environmental modes to describe loss.
This may be chiefly due to absorption in the
phase sample, while noise in the reference arm is tightly controlled
($\gamma_2=0$, $\Gamma^0_2=0$). In gravitational wave detection, closely  losses in both arms
due to imperfections of mirror surfaces, diffraction, and detector
inefficiency are likely\cite{GEO600,romanNatComm}. Fig.~\ref{compare} illustrates the optimal state
for fixed dephasing and increasing values of loss. 

The presence of dephasing ultimately forces the optimal state to take a continuous
form given by a solution to the Schr\"odinger equation \eqref{schrodinger} with potential
\begin{equation}
 \mu(x) = \mu_0 + \tfrac{1}{4} \bigl( \tfrac{r_1}{\frac{1}{2}+x} + \tfrac{r_2}{\frac{1}{2}-x} \bigr)
  \label{muloss}
\end{equation}
that corresponds to placing two \emph{repulsive} Coulomb sources at 
$x=\pm\tfrac{1}{2}$ (see table~\ref{resultstable}). The second source
should be replaced by an infinite
wall in the `one-mode loss' scenario. The constant term is $\mu_0 = N^2 \Gamma^0$,
where $\Gamma^0 = \Gamma^0_1 + \Gamma^0_2$ is the total
dephasing.

For large $N$, the optimal state becomes increasingly localized near
the minimum of \eqref{muloss}. Table~\ref{resultstable} gives phase
estimation error as a sum of classical phase diffusion limit
$\Gamma^0$ and error attributable to loss. Interestingly, errors
attributable to loss in either arm $\epsilon_1/\sqrt{N}$ and
$\epsilon_2/\sqrt{N}$ add in magnitude rather than in quadrature.

A fraction $\frac{\epsilon_1 \epsilon_2} {\epsilon_1^2 - \epsilon_1
  \epsilon_2 + \epsilon_2^2}$ of error in excess of  $\Gamma^0$ in each space of $N_d < N$ total \emph{detected} photons is attributable to `loss-induced dephasing'.
The \emph{total} number of lost particles can be inferred as $N - N_d$, but we cannot learn how the lost photons were distributed between the two interferometer arms. Doing so would erase all phase information in the measurement, we would have projected onto a $S^z$ eigenstate. This uncertainty in the distribution of particles presents a \emph{de facto}
extra source of dephasing.
 
Optimal partitioning of total particle resources $\nu N$ involves interplay between $\Gamma^0$ and the corrections phase variance resulting from higher order terms in the potential. (The contribution  scaling $\propto 1/(\nu N)$ is unaffected by partitioning since the total $\nu N$ is fixed.) These corrections produce novel precision bounds and scaling, indicated in the last row of table II, valid when $r_{1,2} \gg 1$. 

Determining the structure of the optimal states subject to physical decoherence represents an important advance in 
metrology, however the physical generation of optimal states is now a new challenge,
especially for large $N$. Elaborate proposals do exist to `tailor-make' the amplitudes
$\psi_m$ in optical systems \cite{tailored} but it is not clear that these would be 
scaleable. 

On the other hand, we have seen (table II) that in certain noise scenarios the optimal state converges on a Gaussian profile. It has been observed \cite{gauss-profile} that Gaussian-profile probe states may be generated naturally in atom interferometers; they approximate  the ground state of the hamiltonian $S^x - \beta (S^z)^2$. They also occur through the action of a single-axis twisting Hamiltonian on an coherent state. These may represent feasible techniques by which the optimal probes might be realized. 

By learning (i) the ultimate resource/performance trade-off, (ii) the unique form of the associated optimal probes and (iii) possible ways to generate those probes, we are taking three positive steps towards optimal design of next-generation quantum sensors subject to realistic decoherence. 

\section*{Methods}
\begin{footnotesize}
\subsection*{Computing Quantum Fisher Information}
\label{append}
Pure dephasing leads to exponential suppression of off-diagonal
matrix elements
\begin{equation}
  \rho(x,x') = \psi(x) \exp\{-N^2\Gamma^0  (x-x')^2 /2 \} \psi(x'),
  \label{rho}
\end{equation}
where we may view $x=m/N$ as a continuous variable for large $N$
and small dephasing $\Gamma^0$. The Gaussian kernel is a
(imaginary time) Feynman propagator $\ee^{-T}$ for a free particle with
mass $\mu=N^2\Gamma^0$. Writing amplitudes as $\ee^{-V/2}$ with
`potential' $V=-\ln \psi^2(x)$, an operator diagonal in coordinate
representation, we may write \eqref{rho} as a thermal state $e^{-H}$ of an abstract
Hamiltonian $H$. Now $H$ is expressible by a series of commutators of $V$ and $T$ via a Baker-Campbell-Hausdorff (BCH)
identity.\cite{BCH} (More explicit details in Appendix A.)

We apply this novel representation for the Quantum Fisher Information,
valid for an arbitrary unitary shift parameter ($\partial \rho / \partial \theta
= -\ii [U,\rho]$),
\begin{equation}
  F = \bigl< \bigl[ U, 2 \tanh \bigl( \tfrac{1}{2}[H,\bullet] \bigr) U \bigr] \bigr>,
  \label{FU}
\end{equation}
as a series of nested commutators, following  the power series expansion of
the hyperbolic tangent. Here $[H,\bullet] \equiv \ad_H$ represents the
adjoint endomorphism of Lee algebra: $[H,\bullet]U=[H,U]$ and  
$[H,\bullet]^2 U=[H,[H,U]]$ and so on. Angular brackets represent
taking a trace with the density matrix.

Nested commutators in the BCH identity\cite{BCH} and  the expansion \eqref{FU} form a
series in powers of $1/\mu$:
\begin{equation}
  \frac{F}{N^2} = \frac{1}{\mu_0} - \frac{1}{\mu_0^2} \int \psi'^2 \dd x
  + O \biggl( \frac{1}{\mu_0^3} \biggr)
  \label{Fmu}
\end{equation}
retaining the two leading terms and using $U=S^z$ for phase evolution.
In the limit $\mu_0 \gg 1$ this can be rewritten in the more intuitive
form of eqn.~\eqref{Fadd}.

In the more general scenario, when other types of noise are
included, the strength of exponential suppression of
phase-carrying elements of the density matrix varies along the diagonal:
$\exp\bigl[-\tfrac{1}{2}\mu\bigl(\tfrac{x+x'}{2}\bigr)\,(x-x')^2\bigr]$.
Additional suppression $\propto V'' \bigl( \tfrac{x+x'}{2} \bigr)$
results from gradient $\psi'(x)$.

QFI is then given approximately by
\begin{equation}
  \frac{F}{N^2} \approx \int \frac{\psi^2}{\mu(x)+\tfrac{1}{4}V''} \dd x
  \approx \int \biggl[ \frac{\psi^2}{\mu(x)} -
  \frac{\psi'^2}{\mu^2(x)} \biggr] \dd x,
  \label{Fmux}
\end{equation}
where in the latter expression we drop corrections the to coefficient
of $\psi^2(x)$ of orders $O \bigl( \tfrac{1}{\mu^2} \bigr)$ 
and higher (resulting from integration by parts). This new `Lagrangian'-type formulation of QFI  is
generally applicable to any probe state having characteristic length
scale (e.g. width) much larger than $1/\sqrt{\mu}$, a small parameter.

The Euler-Lagrange equation for the variational maximum of \eqref{Fmux}
reduces to `Schr\"odinger' equation \eqref{schrodinger} at the same
approximation level: we use the fact that
$\sqrt{ \langle \mu^2(x) \rangle - \langle \mu(x) \rangle^2} \ll \mu(x)$
[here angular brackets denote integration with $\psi_\text{opt}^2(x)$
corresponding to the ground state of `Schr\"odinger equation].

\subsection*{General Decoherence Model}
\paragraph*{Collective decoherence}
in the most general setting can be described by master equation
\begin{equation}
  \frac{\dd\rho}{\dd\theta} = \frac{\partial\rho}{\partial\theta}
  + \dot\Gamma^0 \frac{\partial\rho}{\partial\Gamma^0}
  + \dot\Gamma^- \frac{\partial\rho}{\partial\Gamma^-}
  +\dot\Gamma^+ \frac{\partial\rho}{\partial\Gamma^+},
\end{equation}
where the first term represents unitary phase evolution 
$\partial\rho/\partial\theta = -\ii \bigl[ S^z, \rho \bigr]$
and $\dot\Gamma^0$, $\dot\Gamma^-$, $\dot\Gamma^+$ are the rates of
collective dephasing, relaxation and excitation, respectively.
Individual processes, e.g. relaxation, are described by
\begin{multline}
  \frac{\partial\rho_{mm'}}{\partial\Gamma^-} =
  f_{m+1} f_{m'+1} \rho_{m+1\,m'+1} - f_m f_{m'} \rho_{mm'} \\
  - \tfrac{1}{2} ( f_m-f_{m'} )^2 \rho_{mm'},
  \label{MG}
\end{multline}
with $f_m=\sqrt{(S-m)(S+m+1)}$.

The first two terms on the right hand side describe
probability-conserving drift and diffusion along the diagonal
$m-m'=\const$. Their effects can be ignored as long $N\Gamma^- \ll 1$. 
The last `absorption' term suppresses off diagonal elements with
exponent that is proportional to $(m-m')^2$ for not too large distance
from the diagonal. Thus relaxation represents as additional dephasing
process, of strength that varies along the main diagonal: 
$N^2 \Gamma^- \tfrac{x^2}{1/4 - x^2}$. The effects of pure dephasing, relaxation and
thermal excitation are combined in a single `potential' $\mu(x)$, presented in table II.

\paragraph*{Individual decoherence} 
does not conserve total spin but the density matrix factorizes into
independent sectors labeled by $S \leqslant \tfrac{N}{2}$. Detailed
analysis presented in Appendix A shows that the master equation
can similarly be viewed as a combination of drift and diffusion,
albeit in two dimensional space with $x=m/N$ and $y=S/N$ as well as
induced dephasing of variable rate
\begin{equation}
  \dot\mu=N\frac{\dot\gamma}{4(y^2-x^2)}
  \qquad (\text{where } 
  \dot\gamma=\dot\gamma^0 + \dot\gamma^- + \dot\gamma^+),
\end{equation}
which must be integrated along the path $\bm{(}x(\theta),y(\theta)\bm{)}$. 
Variable $x$ in QFI functional \eqref{Fmux} represents $x(\theta)$ at
the end of the evolution so that a change of variables would be
required to write it in terms of the original probe state $\psi(x)$.
Equivalently, it can be applied verbatim if $x$ is taken to mean $x(0)$
at the beginning of the evolution.

When integrating the dephasing rate, drift alone may be taken into
account; diffusion results in negligible correction. Using expression
for the drift $(v_x,v_y)$ we observe that
\begin{equation}
  \frac{\dd}{\dd\theta}(y^2-x^2) \equiv 2 y v_y - 2 x v_x = -\dot\gamma(y^2-x^2),
\end{equation}
which is easily integrated. Further integration of the dephasing rate 
$\dot\mu$ starting from point $(x,y)$ at $\theta=0$ produces the
`potential' $\mu(x)$:
\begin{equation}
  \mu(x) = \frac{N(\ee^\gamma-1)}{4(y^2 - x^2)}.
\end{equation}
As expected, minimum error is obtained when the probe state
corresponds to $S=N/2$.

\paragraph*{Particle loss}
is described by the following terms in master equation:
\begin{multline}
  \frac{\partial \rho_{mm'}\super{S}}{\partial \gamma_{1,2}} = 
  f_{S+\frac{1}{2}\,m \pm \frac{1}{2}}^\pm f_{S+\frac{1}{2}\,m' \pm \frac{1}{2}}^\pm 
  \rho_{m \pm \frac{1}{2}\,m' \pm \frac{1}{2}}\super{S+\frac{1}{2}}
  -f_{Sm}^\pm f_{Sm'}^\pm \rho_{mm'}\super{S} \\
  -\tfrac{1}{2}(f_{Sm}^\pm - f_{Sm'}^\pm)^2 \rho_{mm'}\super{S}
  \label{Mloss}
\end{multline}
with $f_{Sm}^\pm = \sqrt{S \pm m}$ and $\gamma_1$,
$\gamma_2$ denoting losses in sample and reference arms
respectively.

As before, we introduce continuous variables $x=m/N$ and $y=S/N$.
The last term in eqn.~\eqref{Mloss} is the rate of induced dephasing
which in the continuous limit approaches $\dot \mu = \frac{N}{4} 
\bigl( \frac{\dot\gamma_1}{y+x} + \frac{\dot\gamma_2}{y-x} \bigr)$.
This must be integrated along the path that solves continuous
equations for the drift, $ \frac{\dd}{\dd\theta} (y \pm x) = -
\dot\gamma_{1,2} (y \pm x)$. Adding a contribution $\mu_0 = N^2
\Gamma^0$ due to collective dephasing we obtain the `potential' of
eqn.~\eqref{muloss} in the main text.

\section*{Optimal Clustering for Collective Dephasing}
Interpolating between upper/lower bounds of eq.\eqref{bounds},  the error in the intermediate regime is 
\begin{align}
\text{var} \theta_{\text{est}} = \frac{\Gamma^0}{\nu} + \frac{\alpha(\mu_0)}{\nu N^2}
\end{align}
where numerical coefficient $\alpha \in [1,\pi^2]$ is itself a function of mass $\mu_0 = \Gamma^0 N^2$. Rewriting in terms of mass: 
\begin{align}
\text{var} \theta_{\text{est}} = \frac{\sqrt{\Gamma^0}}{\nu N} \left( \sqrt{\mu_0} +  \frac{\alpha(\mu_0)}{\sqrt{\mu_0}}\right)
\end{align}
Now if we choose to fix our resources $\nu N$ we can differentiate the bracketed expression with respect to mass to find the minimum $\text{var} \theta_{\text{est}}$. This gives $2 \mu_0 \alpha'(\mu_0) - \alpha(\mu_0) +\mu_0 = 0$, or equivalently $\beta'(\sqrt{\mu_0})+1=0$ for $\beta(\sqrt{\mu_0}) = \alpha(\mu_0)/\sqrt{\mu_0}$. Recasting numerical data for minimum quantum error in the presence of dephasing (generated for multiple $N$ and $\Gamma^0$ values) in terms of $\beta(\sqrt{\mu_0})$ and locating the point at which the gradient $= -1$ unveils the optimal $\mu_0 = 1/2$, see fig.\ref{alphabeta}. Therefore the optimal cluster size is $N_c = 1/\sqrt{2 \Gamma^0}$.
\end{footnotesize}

\section*{Acknowledgements}
GAD acknowledges useful discussions with Roman Schnabel. EHC was supported by the NASA Office of the Chief TechnologistÕs
Space Technology Research Fellowship under the guidance of Dirk Englund and Vadim Smelyanskiy.

\appendix
\onecolumngrid
\section*{\selectfont\Large Appendix A}
\twocolumngrid
\begin{footnotesize}
\subsection*{Semiclassical Expansion of Quantum Fisher Information}
Introducing operator representation with 
\begin{equation}
  V=-\ln \psi^2(x), \qquad T=\frac{1}{2}\ln \frac{2\pi}{\mu} + \frac{P^2}{2\mu},
\end{equation}
continuous limit of the density matrix following
collective dephasing process can be written as
\begin{equation}
  \rho = \ee^{-V/2} \ee^{-T} \ee^{-V/2} = \ee^{-H_0-H_1-\cdots}
\end{equation}
where 
\begin{equation}
\begin{split}
  H_0 &= T+V, \\ %= \tfrac{1}{2} \ln \tfrac{2\pi}{\mu} + \frac{P^2}{2\mu} + V,
  H_1 &= \tfrac{1}{12}[T,[T,V]]+\tfrac{1}{24}[V,[T,V]]
  =-\tfrac{\{P,\{P,V""\}\}}{48\mu^2} + \tfrac{V'^2}{24\mu},
\end{split}
\end{equation}
etc., are successive terms in the Baker-Cambell-Hausdorff expansion
(only odd orders appear for symmetrically-split operators)

Sylvester equation for the symmetric logarithmic derivative of the
density matrix undergoing arbitrary unitary evolution may be written as
\begin{equation}
  \frac{1}{2} \bigl\{ \ee^{-H}, L \bigr\} = -\ii \bigl[ U, \ee^{-H} \bigr],
  \label{sld}
\end{equation}
where we explicitly write the density matrix as a thermal state $\rho=\ee^{-H}$.
Multiplying this by $\ee^{H/2}$ from both left and right 
($\ee^{H/2} \cdots \ee^{H/2} = \ee^{H/2} \cdots \ee^{H/2}$) and
exploiting the representation of $\ee^A B \ee^{-A}$ as
\begin{equation}
  \exp([A,\bullet])B \equiv B + [A,B] + \frac{1}{2!}[A,[A,B]] + \cdots,
\end{equation}
we rewrite eqn.~\eqref{sld} in the following form:
\begin{equation}
  \cosh \bigl( \tfrac{1}{2} [H,\bullet] \bigr) L =
  -2\ii \sinh \bigl( \tfrac{1}{2} [H,\bullet] \bigr) U.
\end{equation}
Since superoperators on both sides are commuting, both sides can be
left-multiplied by $\cosh^{-1} \bigl( \tfrac{1}{2} [H,\bullet] \bigr)$.
[Here $\cosh^{-1} x = 1/\cosh x$, \emph{not} the inverse hyperbolic
cosine denoted $\arcosh x$.] 
Substituting SLD into the expression for the Quantum Fisher Information,
\begin{equation}
  F=\Tr ( \rho L^2 ) = \Tr (\ii [U,L] \rho),
\end{equation}
eqn.~\eqref{FU} of the main text is reproduced.

Expansion of hyperbolic tangent is performed to third order,
\begin{equation}
  F = \langle [U,[H,U]] \rangle - \frac{1}{12} \langle [U,[H,[H,[H,U]]]] \rangle+\cdots;
\end{equation}
using $U=N X$ (where $X \equiv S^z/N$) for phase evolution, leading order
contribution to QFI is
\begin{equation}
  \frac{F_0}{N^2} = \langle [X,[H_0,X]] \rangle = \frac{1}{\mu},
\end{equation}
which represents the classical phase diffusion limit 
$\nu \var \theta_\text{est} = \Gamma^0$. Next order corrections is from
third order terms in BCH identity\cite{BCH} and the expansion of hyperbolic
tangent,
\begin{equation}
  \frac{F_1}{N^2} = \langle [X,[H_1,X]] \rangle - 
  \frac{1}{12} \langle [X,[H_0,[H_0,[H_0,X]]]] \rangle =
  - \Bigl< \frac{V''}{4\mu^2} \Bigr>,
\end{equation}
which is evaluated using integration by parts and noting boundary
conditions $\psi \bigl( \pm \tfrac{1}{2} \bigr) = 0$. These two
leading terms are presented in eqn.~\eqref{Fmu}; higher-order 
commutators form a series expansion in powers of $1/\mu$.
 
\subsection*{Gaussian-profile Probes and Canonical Phase Measurements}
When Gaussian-profile states $\psi(x) \propto \ee^{-K x^2/4}$ are
employed, the  algebra has finite basis: $X^2$ and $P^2$. Closed
form expression for `Hamiltonian' is
\begin{equation}
  H = \tfrac{1}{2} \ln \tfrac{K}{\mu_0} + \arsinh \!\sqrt{\tfrac{K}{4\mu_0}}
  \times\! \Biggl[ \sqrt{K\bigl(\mu_0+\tfrac{K}{4}\bigr)}X^2 
  + \frac{P^2}{\sqrt{K\bigl(\mu_0+\tfrac{K}{4}\bigr)}} \Biggr].
\end{equation}
For the symmetric logarithmic derivative and QFI we obtain, respectively,
\begin{equation}
  \frac{L}{N} = -\ii \frac{P}{\mu_0+K/4}, \qquad 
  \frac{F}{N^2} = \frac{1}{\mu_0+K/4},
\end{equation}
verifying that \eqref{Fadd} is exact.

That the SLD is diagonal in $P$-representation suggests that the optimal 
measurement corresponds to projection onto eigenstates $\ee^{\ii P X}$.
For finite $N$, it must be replaced by a POVM built from overcomplete set of 
phase states $| \theta \rangle = \sum_m \exp(-\ii m \theta) | S, m \rangle$,
so-called `canonical' phase measurement.
For general symmetric pure probe state [$\psi^\ast(x)=\psi(-x)$],
these measurement saturate QFI bound \cite{Braunstein94} and result in
phase-independent distribution of error. For the Cosine probe state of
eqn.~\eqref{opt},
 \begin{equation} \label{pdth}
  p(\delta\theta) = \frac{4}{(N+1)\pi}
  \Biggl( \frac{\sin \frac{\pi}{N+1} 
    \cos \frac{\delta\theta}{2} \cos \frac{\delta\theta}{N+1}}
    {\cos \delta\theta - \cos \frac{\pi}{2j+1}}
  \Biggr)^2
\end{equation}
has variance that is bounded by $\pi^2 / N^2$, approaching that value
asymptotically. In the continuous limit the phase distribution is really the Fourier transform of the input state profile. But the distribution above is not exactly the Fourier transform of a Cosine, which we would expect to be a pair of Delta functions. This is because we are forgetting the `window' function for the state profile, $|x| < 1/2$. This 'top-hat' function would itself transform to a Sinc function and thus we may identify the above phase distribution as a convolution of two Delta functions with a Sinc function.

A further convolution with with the much wider distribution of
the random phase ($\Gamma^0 \gg 1/N^2$) is approximately Gaussian with
increased variance $\Gamma^0 + \var \delta\theta $ thereby saturating the mean-squared-error bound. That it coincides with QFI bound obtained 
independently proves optimality of canonical phase measurements in
this limit.

\subsection*{Analysis of Collective Relaxation/Excitation}
Master equation \eqref{MG} of the main text can be viewed as a
finte-difference approximation to 
\begin{equation}
  \tfrac{1}{N}\,\tfrac{\partial}{\partial \Gamma^-} \rho =
  -\tfrac{\partial}{\partial \bar x} [v(x,x') \rho]
  +\tfrac{\partial^2}{\partial \bar x^2} [D(x,x') \rho]
  - \alpha(x,x') \rho
\end{equation}
with drift, diffusion, and absorption, respectively:
\begin{equation}
v=-f(x)f(x'),\quad D=\tfrac{1}{2N}f(x)f(x'),\quad \alpha=\tfrac{N}{2}[f(x)-f(x')]^2,
\end{equation}
where $f(x)=\sqrt{\frac{1}{4}-x^2}$. In the limit $N \Gamma^- \ll 1$
drift and diffusion can be neglected and the absorption suppresses
off-diagonal elements as 
$exp\bigl[-\mu\bigl( \tfrac{x+x'}{2} \bigr)\frac{(x-x')^2}{2}\bigr]$,
where $\mu(\bar x)=N^2 \Gamma^- f'^2(\bar x)$.

Larger values of relaxation lead to sudden death of
precision. `Potential' $\mu(x)$ is obtained by integrating the
dephasing rate along the path given by the drift (diffusion may be
neglected). Also including dephasing and excitation processes,
\begin{multline}
  \mu(x) = N^2 \Gamma^0 + N^2 \tfrac{\Gamma^- + \Gamma^+}{2}
  [ \sinch s_\ast \cosh (s-s_\ast) -1 ],
\end{multline}
where $s=2 \artanh 2x$ and $s_\ast=N\frac{\Gamma^+ - \Gamma^-}{2}$;
we also use notation $\sinch x = \frac{\sinh x}{x}$. From inequality
\begin{equation}
  1/F \geqslant \min_x \mu(x) = \Gamma^0 + 
  \tfrac{\Gamma^- + \Gamma^+}{2} \bigl[
  \sinch \tfrac{N(\Gamma^- - \Gamma^+)}{2} -1 \bigr]
\end{equation}
we conclude that error increases exponentially once 
$N(\Gamma^- - \Gamma^+) \gtrsim 1$.

\subsection*{Individual Decoherence: Reduced Master Equation}
Although individual decoherence does not conserve total spin, master
equation is invariant under qubit permutation. The density matrix
transforms as a trivial representation of permutation group $\mathcal{S}_N$,
so it must be block-diagonal corresponding to different representations
corresponding to total spin $S$. It admits compressed representation
\begin{equation}
  \rho = \!\sum_{S,m,m'} \rho_{mm'}^{(\!S\!)}
  \!\!\! \sum_{\varpi \in \Pi_S\super{N}} \!\!\!
%  \frac{| m \rangle_\varpi \langle m' |_\varpi}{\bigl| \Pi_S^{(\!N\!)} \bigr|}.
  \frac{\ketsub{m}{\varpi} \!\!\times\!\! \brasub{m'}{\varpi}}{\abs[\big]{\Pi_S\super{N}}}
  \label{rhosym}
\end{equation}
The inner sum runs over the orthonormal basis of $N$-qubit
representation that transforms as spin $S$, having dimension
$\bigl| \Pi_S^{(\!N\!)} \bigr| = 
\tfrac{N!(2S+1)}{\left(\frac{N}{2}+S+1\right)!\left(\frac{N}{2}-S\right)!}$.
For the purposes of computing Quantum Fisher Information we
conveniently ignore the nature of $\rho$ as a $2^N\times2^N$ matrix and
write it as a weighted sum $F=\sum_{S \leqslant \frac{N}{2}} w_S F^{(\!S\!)}$.
Components $F^{(\!S\!)}$ are found by solving the Sylvester
equation for the SLD of reduced density matrix of dimension $2S+1$.

Singling out one of the qubits, the remaining $N-1$ qubits transform
as either spin $S+\tfrac{1}{2}$ or $S-\tfrac{1}{2}$ with relative
weights $W_S^\pm = 
\abs[\Big]{\Pi_{S\pm\frac{1}{2}}\super{N-1}} \Big/ \abs[\big]{\Pi_S\super{N}}$:
\begin{equation*}
  W_S^+ = \frac{(S+1)(N-2S)}{(2S+1)N},\qquad 
  W_S^- = \frac{S[N+2(S+1)]}{(2S+1)N}.
\end{equation*}
From decomposition
\begin{equation}
\begin{split}
  \sum_{\varpi \in \Pi_S^{(\!N\!)}} \ketsub{m}{\varpi} \!\!\times\!\! 
  \brasub{m'}{\varpi} = \hspace{-9em} &\\
  &\sum_{\varpi_+ \in \Pi_{S+\frac{1}{2}}\super{N-1}} \!\!\!\Bigl(
    C_{Sm}^{++} \ket*{\tfrac{1}{2}} \; \ketsub{m-\tfrac{1}{2}}{\varpi_+} 
    \!\!\!+C_{Sm}^{-+} \ket*{-\tfrac{1}{2}} \; \ketsub{m+\tfrac{1}{2}}{\varpi_+}
  \!\Bigr) \\[-3ex] & \hspace{5em} \times\! \Bigl(
    C_{Sm'}^{++} \bra*{\tfrac{1}{2}}\brasub{m'-\tfrac{1}{2}}{\varpi_+\!}
    +C_{Sm'}^{-+} \bra*{-\tfrac{1}{2}}\brasub{m'+\tfrac{1}{2}}{\varpi_+\!}
    \Bigr) \\[1ex]
    +\!\!&\sum_{\varpi_- \in \Pi_{S-\frac{1}{2}}\super{N-1}} \!\!\!\Bigl(
      C_{Sm}^{+-} \ket*{\tfrac{1}{2}} \; \ketsub{m-\tfrac{1}{2}}{\varpi_-}
      \!\!\!+C_{Sm}^{--} \ket*{-\tfrac{1}{2}} \; \ketsub{m+\tfrac{1}{2}}{\varpi_-}
      \!\Bigr) \\[-3ex] & \hspace{5em} \times\! \Bigl(
        C_{Sm'}^{+-} \bra*{\tfrac{1}{2}} \brasub{m'-\tfrac{1}{2}}{\varpi_-\!}
        +C_{Sm'}^{--} \bra*{-\tfrac{1}{2}} \brasub{m'+\tfrac{1}{2}}{\varpi_-\!}
      \Bigr)
\end{split}
\label{clebsch}
\end{equation}
with Clebsch-Gordan coefficients
$C_{Sm}^{\pm+} = \pm \sqrt{\tfrac{S+1 \mp m}{2(S+1)}}$,
$C_{Sm}^{\pm-} = \sqrt{\tfrac{S \pm m}{2S}}$,
 we may be able to see that Lindblad operator acting on an isolated qubit,
$\mathcal{L}_k = \pi_k \rho \pi_k^\dag - 
\tfrac{1}{2} \bigl\{ \pi_k^\dag \pi_k, \rho \bigr\}$ (where $\pi_k$
can be $s_k^z$, $s_k^-$, or $s_k^+$)
has non-zero projections on states of spin $S \pm 1$ as well as $S$.
Differential change of the reduced density matrix may be found by
taking a trace with projection operators
$\sum_{\varpi} \ketsub{m}{\varpi} \!\!\times\!\! \brasub{m'}{\varpi}$ 
with $\varpi \in \Pi_{S\pm1}\super{N}$ or $\varpi \in \Pi_S\super{N}$
respectively; eventual summation over all contributions ($k=1,\ldots,N$) 
restores the symmetric form \eqref{rhosym}.

Using this approach, reduced master equation for the combined effect
of individual dephasing, relaxation and excitation can be written in
compact form
\begin{multline}
  \frac{1}{N}\frac{\partial \rho_{mm'}\super{S}}{\partial \gamma^\sigma} = 
   \sum_{\substack{\tau=\pm \\ \delta=0,1}} W_{S-\tau\delta} ^\tau 
  f_{S-\tau\delta\,m-\sigma}^{\sigma\tau\delta} f_{S-\tau\delta\,m'-\sigma}^{\sigma\tau\delta} 
  \rho_{m-\sigma\,m'-\sigma}\super{S-\tau\delta}\\
  -\rho_{mm'}\super{S} \!\! \sum_{\substack{\tau=\pm \\ \delta=0,1}}
  W_S^\tau f_{Sm}^{\sigma\tau\delta} f_{Sm'}^{\sigma\tau\delta}
  -\rho_{mm'}\super{S} \!\! \sum_{\substack{\tau=\pm \\ \delta=0,1}}
  W_S^\tau \bigl( f_{Sm}^{\sigma\tau\delta} -
  f_{Sm'}^{\sigma\tau\delta} \bigr)^2 \!\!\big/ 2\\[-2.5ex]
\label{rhomms}
\end{multline}
with coefficients $f_{Sm}^{0\tau\delta} = \Bigl( 
  C_{Sm}^{+\,\tau} C_{S+\tau\delta\,m}^{+\:\cramped{\tau(\!-1\!)^\delta}}\!\!
  -C_{Sm}^{-\,\tau} C_{S+\tau\delta\,m}^{-\:\cramped{\tau(\!-1\!)^\delta}} \Bigr)\Big/2$
and $f_{Sm}^{\pm\tau\delta} = C_{Sm}^{\mp\,\tau}
C_{S+\tau\delta\:m\pm1}^{\pm\,\cramped{\tau(\!-1\!)^\delta}}$.

The integral of motion for this equation is $m-m'=\const$, as
expected. The third term represents `absorption',
\begin{equation}
\begin{split}
  \alpha &= \sum_{\sigma,\tau,\delta} W_S^\tau \bigl( 
  f_{Sm}^{\sigma\tau\delta} - f_{Sm'}^{\sigma\tau\delta} \bigr)^2\big/2, \\
  \text{while }\bigl(\begin{smallmatrix} v_x \\ v_y \end{smallmatrix}\bigr) &=
  \sum_{\sigma,\tau,\delta} 
  \bigl(\begin{smallmatrix} \sigma \\ \tau\delta \end{smallmatrix}\bigr)
  W_S^\tau f_{Sm}^{\sigma\tau\delta} f_{Sm'}^{\sigma\tau\delta} \dot\gamma_\sigma \\
  \text{and } \Bigl(\begin{smallmatrix} D_{xx} & D_{xy} \\ 
      D_{yx} & D_{yy} \end{smallmatrix}\Bigr) &=
  \frac{1}{N} \sum_{\sigma,\tau,\delta}
  \bigl(\begin{smallmatrix} \sigma^2 & \sigma\tau\delta \\ 
    \sigma\tau\delta & \delta \end{smallmatrix}\bigr)
  W_S^\tau f_{Sm}^{\sigma\tau\delta} f_{Sm'}^{\sigma\tau\delta} \dot\gamma_\sigma,
\end{split}
\end{equation}
are `drift' and `diffusion' along the diagonals from the first two
probability conserving terms in eqn.~\eqref{rhomms}. The `absorption'
terms suppresses off-diagonal elements, resulting in effective rate
of dephasing $\dot \mu$. In a continuous limit
\begin{equation}
\begin{split}
  \dot \mu &= N \tfrac{\dot\gamma^0+\dot\gamma^-+\dot\gamma^+}{4(y^2-x^2)}, \\
  v_x &= -\bigl(\tfrac{1}{2}+x\bigr)\dot\gamma^- + 
  \bigl(\tfrac{1}{2}-x\bigr)\dot\gamma^-,\\
  v_y &= -\tfrac{y^2-x^2}{2y} \dot\gamma^0 
  -\tfrac{y^2+x(1+x)}{2y} \dot\gamma^-
  -\tfrac{y^2-x(1-x)}{2y} \dot\gamma^+, \\
\shortintertext{and, less importantly, diffusion coefficients}
  D_{xx}&=\tfrac{1}{N}\bigl[\bigl(\tfrac{1}{2}+x\bigr)\dot\gamma^- 
  + \bigl(\tfrac{1}{2}-x\bigr)\dot\gamma^-\bigr]\\
  D_{xy}&=\tfrac{1}{N}\bigl[\tfrac{y^2+x(1+x)}{2y} \dot\gamma^-
  -\tfrac{y^2-x(1-x)}{2y} \dot\gamma^+ \bigl]\\
  D_{yy}&=\tfrac{1}{N}\bigl[\tfrac{y^2-x^2}{4y^2} \dot\gamma^0
  +\bigl(\tfrac{y^2+x^2}{4y^2}+x\bigr) \dot\gamma^-
  +\bigl(\tfrac{y^2+x^2}{4y^2}-x\bigr) \dot\gamma^+\bigr].
\end{split}
\end{equation}
The dephasing rate $\dot \mu$ is integrated along the trajectory
described by drift as presented in Methods.

\end{footnotesize}


\begin{thebibliography}{99}

\bibitem{Giovanetti04}
V.~Giovanetti, S.~Lloyd, and L.~Maccone,
Science {\bf 306}, 1330 (2004).

\bibitem{Giovanetti11}
V. Giovannetti, S. Lloyd, and L. Maccone, 
Nature Photon. {\bf 5}, 222 (2011).


\bibitem{knysh}
S.~Knysh, V.~N.~Smelyanskiy, and G.~A.~Durkin,
Phys. Rev. A {\bf 83}, 021804(R) (2011).

\bibitem{RDD12}
R. Demkowicz-Dobrzanski, J. Kolodynski, and M. Guta,
Nature Comm. {\bf 3}, 1063 (2012).

\bibitem{diffother} B.~M.~Escher, R.~L.~de~Matos Filho, K.~Davidovich,
Nature Physics {\bf 7}, 406 (2011).

\bibitem{fujiwara} A.~Fujiwara and H.~Imai,
J. Phys. A {\bf 41}, 255304 (2008).

\bibitem{Escher12}
B. M. Escher, L. Davidovich, N. Zagury, and R. L. de Matos Filho, 
Phys. Rev. Lett. {\bf 109}, 190404 (2012).

\bibitem{Paris11}
M. G. Genoni, S. Olivares, M. G. A. Paris
Phys. Rev. Lett {\bf 106}, 153603 (2011).

\bibitem{NOON}
B. C. Saunders,
Phys. Rev. A { \bf 40 }, 2417�2427 (1989);
Dowling, J. P.,
Contemporary Physics, { \bf 49 } (2), 125-143 (2008).

\bibitem{NOON-crap}
B. R. Bardhan,, K. Jiang1, and J. P. Dowling,
Phys. Rev. A { \bf 88}, 023857 (2013).

\bibitem{Braunstein94}
S.~L.~Braunstein and C.M.~Caves,
Phys. Rev. Lett. {\bf 72}, 3439 (1994).

\bibitem{ParisQmTech}
M. G. A. Paris,
Int. J. Quant. Inf. {\bf 7}, 125 (2009).

\bibitem{mastereqn}
D.F. Walls and G.J. Milburn,
\emph{Quantum Optics}, 
Springer, Berlin (1996).


\bibitem{Nori}
J. Ma, X. Wang, C. P. Sun and F. Nori,
Phys, Rep.  {\bf 509}, 89-165 (2011).

\bibitem{DepolChan}
M. Sasaki, M. Ban, and S. M. Barnett,
Phys. Rev. A  {\bf 66}, 022308 (2002).

\bibitem{BECMZ}
C. Lee, 
Phys. Rev. Lett. {\bf 97}, 150402 (2006).

\bibitem{Yurke}
B. Yurke, S. L. McCall, and J. R. Klauder,
Phys. Rev. A  {\bf 33}, 4033 (1986).

\bibitem{clocks}
S. A. Diddams, J. C. Bergquist, S. R. Jefferts, and C. W.
Oates, Science { \bf 306}, 1318 (2004).

\bibitem{GHZ}
J. J. Bollinger, W. M. Itano, D. J. Wineland, and D. J.
Heinzen, Phys. Rev. A { \bf 54}, R4649 (1996).

\bibitem{dephasing10}
Y. C. Liu, G. R. Jin, and L. You,
Phys. Rev. A {\bf 82}, 045601(2010).

\bibitem{Roman06} A. Franzen, B.  Hage, J. DiGuglielmo,
J. Fiur\'{a}\v{s}ek, and R. Schnabel,
Phys. Rev. Lett. \textbf{97}, 150505 (2006).

\bibitem{Caves80} C.~M.~Caves, Phys. Rev. Lett. {\bf 45}, 75 (1980).

\bibitem{fibershapesensor}
V. G. M. Annamdas,
Int. J. Mat. Eng.; {\bf 1} (1): 1-16  (2011)

\bibitem{Blatt14Qubits}
T. Monz, P. Schindler, J. T. Barreiro, et al., 
Phys. Rev. Lett. { \bf 106}, 130506 (2011).

\bibitem{PeggSummy} 
G.S. Summy and D.T. Pegg,
Optics Comm. {\bf 77}, 1. 75-79, (1990); 
D. W. Berry and H. M. Wiseman,
Phys. Rev. Lett. {\bf 85}, 5098 (2000).

\bibitem{Josh}
J. Combes and H. M. Wiseman
J. Opt. B: Quantum Semiclass. Opt. { \bf 7} 14-21 (2005) 

\bibitem{spin-sq}
M. Kitagawa and M. Ueda, 
Phys. Rev. A { \bf 47}, 5138 (1993).

\bibitem{LukinSpin}
A Andre and M.D. Lukin,
Phys. Rev. A, { \bf 65} 5, 053819 (2002).

\bibitem{BECoptimal}
G. R. Jin, Y. An, T. Yan, and Z. S. Lu,
Phys. Rev. A { \bf 82}, 063622 (2010).

\bibitem{vourdas}
A. Vourdas, Phys. Rev. A {\bf 41}, 1653 (1990).

 
\bibitem{supplement}
See Appendix A.

\bibitem{Caves81}
C. M. Caves, Phys. Rev. D {\bf 23}, 1693 (1981).


\bibitem{Holland93}
M.~J.~Holland and K.~Burnett,
Phys. Rev. Lett {\bf 71}, 1355 (1993).

\bibitem{wigner} Wigner rotation element $d^{S}_{m,m'}(\chi) = \langle S, m |\exp \{ - i \chi \hat{S}^y \} | S, m' \rangle$.

\bibitem{phase-meas}
S. M. Barnett,  
Philosophical Transactions of the Royal Society of London. Series A: Mathematical, Physical and Engineering Sciences { \bf 355} 1733 (1997): 2279-2290


\bibitem{UncertPhase}
S. M. Barnett and D. T. Pegg,
Phys. Rev. A {\bf 41}, 3427 (1990).

\bibitem{Dicke}
R. H. Dicke,
Phys, Rev. {\bf 93}, 99-110 (1954).

\bibitem{Huelga1997}
S. F. Huelga, C. Macchiavello, T. Pellizzari, et al.,
Phys. Rev. Lett. {\bf 79}, 3865 (1997).

\bibitem{local-decoh}
M. Foss-Feig, K. R. A. Hazzard, J J Bollinger, A. M. Rey, and C W Clark,
New J. Phys. { \bf 15} 113008, (2013).

\bibitem{BECloss}
Yun Li, P. Treutlein, J. Reichel and A. Sinatra,
Euro. Phys. J. B { \bf 68},  365 (2009); 
S. Knoop, J. S. Borbely, R. van Rooij, and W. Vassen,
Phys. Rev. A { \bf 85}, 025602 (2012).
 
\bibitem{fn4}
This is Kummer confluent hypergeometric function: $_1M_1(a,b;z) = 1 +
\frac{a}{b} z + \frac{a(a+1)}{b(b+1)} \frac{z^2}{2!}+ \dots
$

\bibitem{GEO600}
J. Abadie, et al.,
Nature Physics, { \bf 7} (12). pp. 962-965, (2011)

\bibitem{romanNatComm}
R. Schnabel, N. Mavalvala, D.E.  McClelland, and P.K.Lam, 
 Nat. Comm, { \bf 1}, 121.

\bibitem{tailored}
M. Dakna, J. Clausen, L. Kn\"{o}ll, and D.-G. Welsch,
Phys. Rev. A { \bf 59}, 1658 (1999).

\bibitem{gauss-profile} 
I.Tikhonenkov, M. G. Moore, and A. Vardi
Phys. Rev. A { \bf 82}, 043624 (2010).

\bibitem{BCH}
N. Hatano and M. Suzuki, 
\emph{Finding exponential product formulas of higher orders.} 
Quantum Annealing and Other Optimization Methods. 
Springer Berlin, 37-68 (2005).

\end{thebibliography}
\end{document}